# Picophotonics - Subatomic Optical Localization Beyond Thermal Fluctuations

Tongjun Liu[1]*, Cheng-Hung Chi[1], Jun-Yu Ou[1], Jie Xu[1], Eng Aik Chan[2], Kevin F. MacDonald[1], and Nikolay I. Zheludev[1, 2]

[1]Optoelectronics Research Centre and Centre for Photonic Metamaterials, University of Southampton; Highfield, Southampton, SO17 1BJ, UK.

[2]Centre for Disruptive Photonic Technologies, School of Physical and Mathematical Sciences and The Photonics Institute, Nanyang Technological University, Singapore; 637378, Singapore.

Despite recent tremendous progress in optical imaging and metrology, the resolution gap between atomic scale transmission electron microscopy and optical techniques has not been closed. Is optical imaging and metrology of nanostructures exhibiting Brownian motion possible with resolution beyond thermal fluctuations? Here we report on an experiment in which the average position of a nanowire with a thermal oscillation amplitude of ~150 pm is resolved in single-shot measurements with precision of 92 pm using light at a wavelength of $\lambda$ = 488 nm, providing the first example of such sub-Brownian metrology with ~$\lambda$/5,300 precision. To localize the nanowire, we employ a deep learning analysis of the scattering of topologically structured light, which is highly sensitive to the nanowire's position. As a non-invasive optical metrology with sub-Brownian absolute errors, down to a fraction of the typical size of an atom (Si: 220 pm diameter), it opens the exciting field of picophotonics.

Over the past decade, spatial resolution in far-field optical imaging and metrology has improved far beyond the classical Abbe diffraction limit of $\lambda$/2, where $\lambda$ is the wavelength of light. A variety of fluorescence- and structured illumination-based, deterministic and stochastic (optical fluctuation and single-molecule localization), microscopy techniques[1-7], now commonly used in biological imaging, routinely achieve resolution of a few tens of nanometers, or better than $\lambda$/10. The application of artificial intelligence to the analysis of coherent light scattered by an object offers metrology with an accuracy of only a few nanometers[8], or better than $\lambda$/100, on a par with scanning electron microscopy. In what follows, we demonstrate an approach to optical measurements with uncertainty reaching a level of $\lambda$/5,300 – a "sub-Brownian" length scale, equivalent to a fraction of the typical size of an atom, and shorter than the thermal motion amplitude of the target objects. In comparison with interferometric techniques, which can provide high sensitivity to changes in optical path length (as opposed to lateral displacement) - down to ~$10^{-19}$ m in the ultimate incarnation of gravitational wave detectors with km-scale baselines[9], our approach allows for single-shot measurements of a *micro- to nanoscopic object* through a deep leaning-enabled analysis of its scattering pattern when it is illuminated with coherent, topologically structured light containing deeply subwavelength singularity features.

In experiment, we measure the in-plane position of a suspended 17 μm long, 200 nm wide nanowire, cut by focused ion beam milling from a 50 nm thick $Si_3N_4$ membrane coated with 65 nm gold, relative to a fixed edge of the surrounding membrane with a 100 nm gap on either side (Fig. 1). This position, i.e. the displacement of the nanowire from the central position between two edges of the slit, can be controlled electrostatically with high accuracy over a few nanometer range through the application of a DC bias across the gap. The sample was illuminated by a coherent light at a wavelength of $\lambda$ = 488 nm, with either a plane (defocused Gaussian) wavefront or a superoscillatory wavefront profile formed by a spatial light modulator-based wavefront synthesizer (see

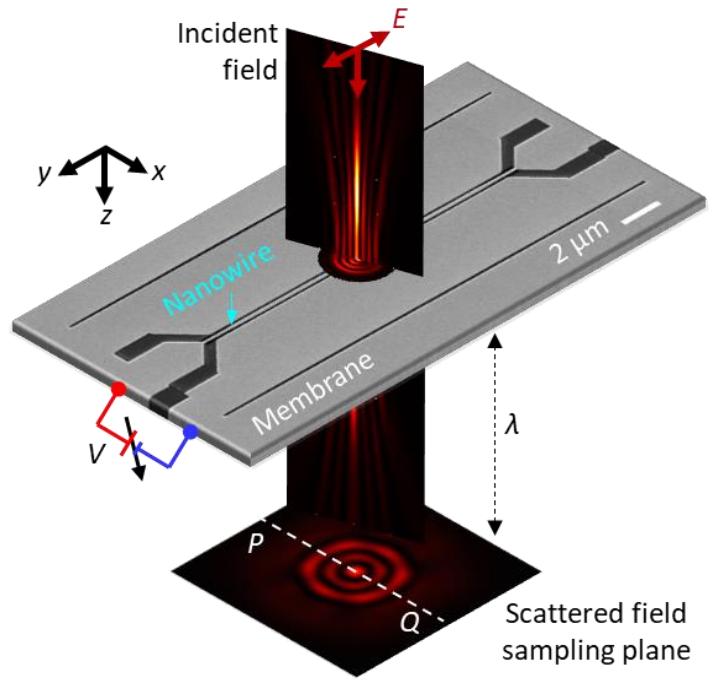

**Fig. 1. Measuring nanowire displacement via scattering of topologically structured light**. Incident light scattered from the nanowire is imaged in transmission through a high-NA microscope objective (not shown) focused in a plane located at a distance $\lambda$ behind the membrane. Deeply-subwavelength lateral (*x*-direction) displacements of the wire, controlled by application of a DC bias between the wire and the adjacent edge of the supporting membrane, are quantified via a deep-leaning enabled analysis of single-shot scattering patterns.

Supplementary Materials section S1). The intensity pattern of light scattered by the nanostructure was imaged in transmission at a distance of $\sim\lambda$ (~0.5 µm) from the membrane by a 16-bit image sensor through a microscope objective with a numerical aperture of 0.9.

To enable optical measurements of unknown nanowire positions, we created a dataset of single-shot (401 pixel × 401 pixel) scattering patterns recorded at 301 different (electrostatically controlled) positions of the nanowire in random sequence. (To exclude any effect of stress history in the nanowire, its position was reset to zero every between each measured position.) Knowledge of the nanowire position is obtained from *a priori* measurements, under a scanning electron microscope, of the dependence of nanowire position on applied bias. Eighty percent of these images, selected at random, were used for neural network training, i.e., as scattering patterns for known positions of the nanowire. (Detail of the network architecture and training procedure, and positional calibration measurements are given in Supplementary Materials section S2.) The trained network was then tasked with determining (nominally) unknown nanowire positions from previously unseen single-shot scattering patterns. This exercise was conducted for regimes of plane wave and superoscillatory illumination.

Figure 2 shows the results of such measurements, in terms of the correlation between optical measurements of nanowire position, retrieved by the trained neural network from scattering patterns, and ground truth (*a priori* calibrated) displacement values for plane wave (Fig. 2a) and superoscillatory (Fig. 2b) illumination. The statistical spread of datapoints is derived from twenty independent neural network training and testing cycles, each yielding one measurement at sixty randomly selected actual displacement values [see Supplementary Information section S2]. We characterise performance of the technique in terms of optical measurement precision and accuracy:

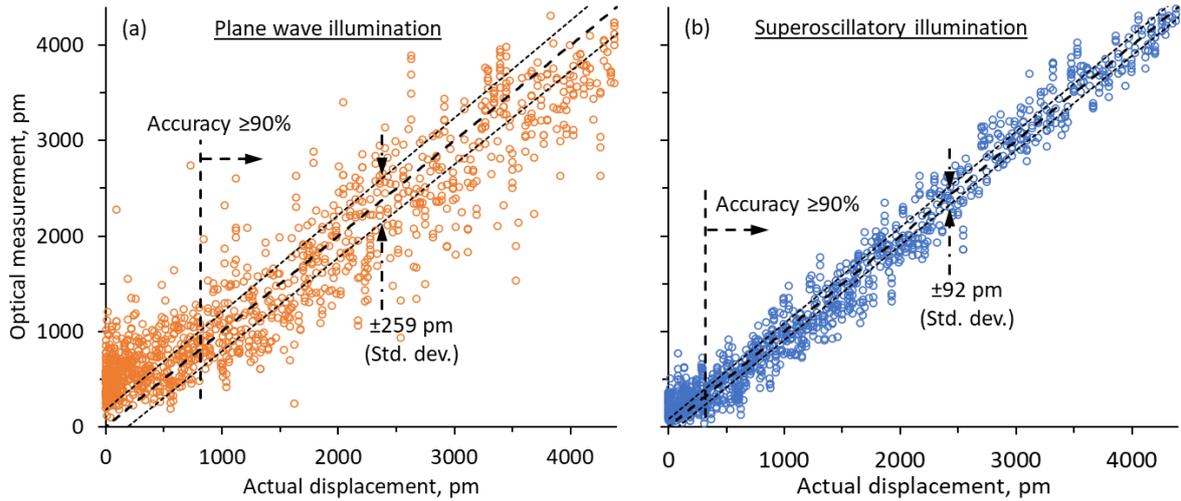

**Fig. 2. Optical measurements of nanowire displacement.** Optically measured versus actual values of nanowire displacement for (a) plane wave and (b) topologically structured superoscillatory illumination. Dotted lines above and below the ideal correlation diagonals are plotted at ±1 standard deviation. Dashed vertical lines denote minimum nanowire displacements that can be measured with accuracy exceeding 90% (relative error <10%).

precision being how close repeated measurements at a given value of actual displacement are to one another (evaluated in terms of measurement standard deviation, averaged over the entire range of measured values); accuracy being how close measurements are to the true value (evaluated in terms of relative error – difference between measured and true values as a percentage of the true value).

Our results show that nanowire displacements can be measured with precision (mean standard deviation – Fig. 2b) of 92 pm with superoscillatory illumination, and 259 pm using plane wave illumination. At a wavelength of 488 nm, a precision of 92 pm represents a level of $\lambda/5,300$, and should be compared with the 775 pm lattice parameter of silicon nitride (the material from which the nanowires are manufactured) and the ~220 pm covalent diameter of a silicon atom. With superoscillatory illumination, accurate measurements are possible for displacements down to a few hundred picometres, with relative error ≤10% down to 320 pm; For plane wave illumination this accuracy threshold is at 816 pm.

Optical metrology based upon analysis of scattered light is an inverse problem that can be reduced to the Fredholm integral equation, which can be efficiently solved by a neural network[8]. Recent work has demonstrated that this approach yields accuracy better than $\lambda/100$ in measuring the width of gaps in an opaque film with plane wave illumination, using a neural network trained on a set of nanofabricated samples with a range of different gap sizes. There are two major contributing factors to the hundredfold improvement in precision reported here: a markedly better training process and the use of topologically structured superoscillatory light.

Precisely tailored interference of multiple waves can form intensity "hotspots" in free space, with dimensions smaller than the conventional diffraction limit, as a manifestation of what is known as superoscillation[10]. The electromagnetic field near a superoscillatory hotspot has many features similar to those in the vicinity of resonant plasmonic nanoparticles or nanoholes - hotspots are surrounded by phase singularities and nanoscale zones of energy backflow where phase gradients can be more than an order of magnitude larger than in a free propagating plane waves[11].

The use of such topologically structured light gives an advantage for AI-enabled metrology: The ability to evaluate small changes in the position of the nanowire depends upon the magnitude of associated changes in the scattered light field at distance $z$ from the object $A(x, z)e^{i\phi(x)} =$ $f(A_0(x,0)e^{i\phi_0(x,0)})$, where $A_0(x,0)$ is the amplitude and $\phi_0(x,o)$ is the phase of the incident light in the *xy* object plane. A small displacement in the object against the incident field in the *x*-direction results in a change in scattered light intensity $\delta I(x) \sim \delta A_0(x,0)^2 + A_0(x,0)^2 \delta\phi_0(x,0)^2$. The first term in this expression is related to the change of illumination intensity associated with the object's positional shift, while the second relates to the corresponding change in phase. The phase-dependent term is absent for plane wave illumination, but can be large under superoscillatory illumination, when the object traverses a small (deeply subwavelength) feature of the incident field, such as singularity, where the phase $\phi_0(x,0)$ jumps by $\pi$.

The responses of scattered plane wave and topologically structured light fields to displacement of an illuminated nanowire are illustrated, through computational modelling, in Fig. 3. The incident superoscillatory wavefront (detailed in Ref. [11]) has a central intensity maximum (Fig. 3a) flanked by phase singularities and zones of high phase gradient (Fig. 3b). We consider the case here where these singularities lie in the nanowire sample plane. As a figure of merit for the sensitivity of the scattered field to small displacements of the nanowire, Fig. 3c presents the magnitude of the relative change in scattered light intensity induced by a $\lambda/1000$ (~0.5 nm) shift in nanowire position, as a function of (horizontally) image plane coordinate and (vertically) the initial position of the sample within the structured light field. The scattered field intensity is strongly dependent on both, with largest changes (of up to 0.1%), occurring when a sharp edge of the nanostructure coincides with a phase singularity in the incident superoscillatory field. For comparison, Fig. 3d shows the same for plane wave illumination. Here, the variations in scattered field intensity are smaller (reaching only 0.03%) and relatively weakly dependent on both image plane coordinate and lateral position of the sample. The contrast between Figs. 3c and 3d explains the better precision of positional measurement achieved with superoscillatory, as compared to plane wave, illumination (Fig 2).

The quality of artificial intelligence is directly related to the quality of training data for the neural network. Our ability here to achieve picometric levels of precision results firstly from the use of a training set that is ultimately congruent with the object of interest: the same electrostatically reconfigurable nanostructure is used for training and as the object of metrological study. Moreover, ground truth positional displacement values are more precisely calibrated in this singular electrostatically controlled gap sample (Fig. S1) than they are in previously employed sets of mutually independent training samples fabricated with typically few-nanometer tolerance[8, 12].

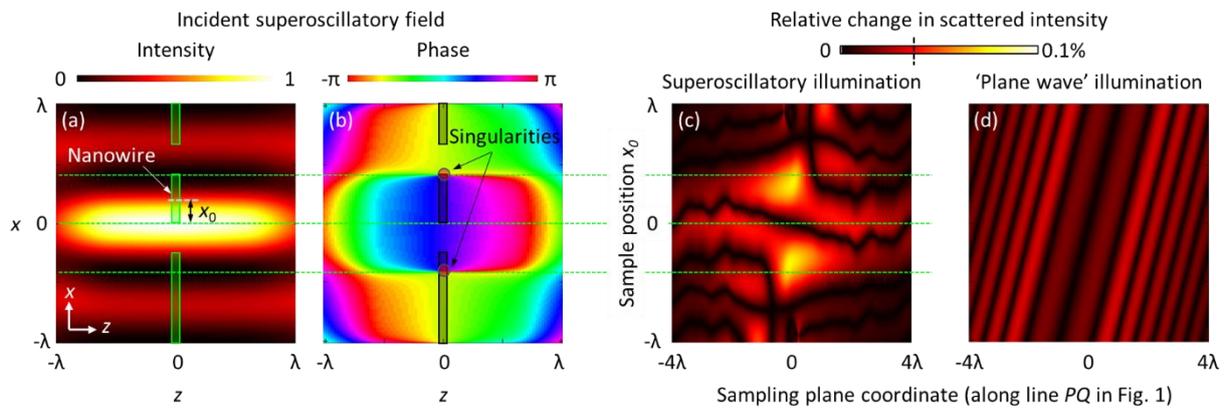

**Fig. 3. Sensitivity of scattered fields to small nanowire displacements.** (a) Intensity and (b) phase profiles of the superoscillatory field in the *xz* plane [light propagating in the *+z* direction, wavelength $\lambda$ = 488 nm]. The sample – a nanowire in the gap between two semi-infinite sections of membrane – lies in the *z* = 0 plane [its cross-sectional profile being shown in green in (a) and grey in (b)]. (c) Relative change in scattered light intensity resulting from a $\lambda/1000$ displacement of the sample in *x*-direction along a cross-sectional line through the scattering pattern in the sampling plane [the line *PQ* in Fig. 1] as a function of the initial position $x_0$ of the sample relative to the symmetry axis of the light field. (d) Corresponding plot of relative change in scattered light intensity for plane wave illumination of the same sample structure. [Further detail of numerical simulations is given in Supplementary information section S3.]

It should further be noted that access to this low level of measurement error is only possible because deep learning-based retrieval process is sensitive primarily to the intensity profile of the diffraction pattern, but less so to its lateral position. As such it is strongly resilient to small thermal and mechanical drifts in the mutual positions of the lens, sample, and incident light beam during the experiment[12].

Our results represent the first example of "sub-Brownian" optical metrology – the measurement of object dimensions/displacements with precision smaller than the amplitude of its thermal motion. The amplitude of nanowire thermal vibration can be evaluated from the Langevin oscillator model[13]: in the present case, the nanowire's fundamental in- and out-of-plane oscillatory modes, at 1.6 and 1.1 MHz respectively, have amplitudes of 145 and 215 pm at room temperature (see supplementary materials S4), i.e. values larger than the achieved superoscillatory illumination measurement precision of 92 pm. This is possible because our measurements are performed with a detector integration time of ~100 ms and thus return the mean position of the nanowire, which oscillates thermally with a much shorter (~0.6 μs) period. Measurements are single-shot, and do not require scanning of the object, so they can be performed in binning mode, with a frame rate equal to that of the image sensor which may reach tens of megahertz in today's ultrafast cameras. It should be noted here that the effect of optical forces on nanowire position is negligible (see Supplementary Materials section S5).

AI-enabled analysis of scattering in topologically structured light fields offers access, under ambient conditions, to levels of resolution in optical localization measurement that are otherwise only attainable in scanning tunnelling microscopy. Localization precision surpassing the diffraction limit of conventional microscopes thousands of times over has been demonstrated here on a system allowing for the collection of *in-situ* physical training data for the neural network. While a dependence on *in-situ* training may be seen as a limitation, the approach can be applied in a variety of systems where a regime of externally controlled positioning is available for training, to then enable non-invasive study of motion such as may be induced by ambient forces and fields (internal or external to the object), or thermal motion. This may be, for example in: non-contact position monitoring of platforms in ultra-precise STM/AFM instruments; monitoring positional displacements in MEMS and NEMS devices such as accelerometers and nanomachines; monitoring structural deformations and thermal drifts in precise instruments (e.g. telescope mirrors) and seismographs; measuring the thermal expansions of macroscopic objects and monitoring nano-gaps affected by microkelvin temperature variations. Moreover, there is growing interest in short-range forces and associated phenomena at the nano- to picoscale, which may be investigated via comparative studies in in-situ trained systems designed to enhance or suppress the mechanism of interest, or where external stimuli (e.g. light) can be selectively applied or withdrawn. Forces of interest may include, for example: ponderomotive optical, Van der Waals, Casimir, and recently identified non-Hamiltonian and spin-related optical forces[14, 15], which manifest at nano- to picometer scales. Indeed, while our technique cannot compete with LIGO-type macroscopic platform displacement measurements on kilometric baselines for gravitational wave detection[9], it may be useful in the study of micro-gravitational forces on micro/nano-objects at very short distances[16]. With ultrafast image sensors (>10 Mfps, as are now becoming accessible), our method may be applied to the study of dynamics and transient processes at the (sub)nanoscale. For example: the study of Brownian motion thermodynamics of nano-objects, including the ballistic regime[17] and non-Markovian processes of thermal fluctuation, which may have applications in fast thermometry and mass measurement; the study of electron and plasmon quantum transport through atomic scale gaps[18-20]; configuration chemistry of individual molecules; protein folding[21]; and other time-dependent events in macromolecules, nanomachines and 2D materials, where flexural, phononic modes are now understood to be of critical importance to thermal, electrical and mechanical properties[22-24].

Optimization of the incident field topology, beyond the simple superoscillatory wavefront used in this study, is expected to allow for even higher measurement precision in comparison to plane wave illumination. As for any other optical visualization technique, performance of the reported method depends upon the refractive index contrast between the object of interest and its surroundings and, as such, it may be efficiently applied to localization studies in metallic, high-index dielectric and semiconductor nanostructures and metamaterials.

**Acknowledgements**

This work was supported by the Engineering and Physical Sciences Research Council, UK (grant numbers EP/M009122/1 and EP/T02643X/1; NIZ, KFM, JYO), the Ministry of Education, Singapore (MOE2016-T3-1-006; NIZ), and the China Scholarship Council (201806160012; TL).

**References**

1. Hell S. W. & Wichmann J. Breaking the diffraction resolution limit by stimulated emission: stimulated-emission-depletion fluorescence microscopy. *Opt. Lett.* **19**, 780-782 (1994).
2. Rust M. J., Bates M. & Zhuang X. Sub-diffraction-limit imaging by stochastic optical reconstruction microscopy (STORM). *Nat. Methods* **3**, 793-796 (2006).
3. Betzig E., *et al.* Imaging intracellular fluorescent proteins at nanometer resolution. *Science* **313**, 1642-1645 (2006).
4. Hess S. T., Girirajan T. P. & Mason M. D. Ultra-high resolution imaging by fluorescence photoactivation localization microscopy. *Biophys. J.* **91**, 4258-4272 (2006).
5. Guerra J. M. Super-resolution through illumination by diffraction-born evanescent waves. *Appl. Phys. Lett.* **66**, 3555-3557 (1995).
6. Gustafsson M. G. L. Nonlinear structured-illumination microscopy: Wide-field fluorescence imaging with theoretically unlimited resolution. *Proc. Natl. Acad. Sci.* **102**, 13081-13086 (2005).
7. Super-resolution microscopy: https://en.wikipedia.org/wiki/Super-resolution_microscopy.
8. Rendón-Barraza C., *et al.* Deeply sub-wavelength non contact optical metrology of sub-wavelength objects. *APL Photon.* **6**, 066107 (2021).
9. Cahillane C. & Mansell G. Review of the Advanced LIGO Gravitational Wave Observatories Leading to Observing Run Four. *Galaxies* **10**, 36 (2022).
10. Zheludev N. I. & Yuan G. Optical superoscillation technologies beyond the diffraction limit. *Nat. Rev. Phys.* **4**, 16-32 (2022).
11. Yuan G., Rogers E. T. F. & Zheludev N. I. "Plasmonics" in free space: observation of giant wavevectors, vortices, and energy backflow in superoscillatory optical fields. *Light Sci. Appl.* **8**, 2 (2019).
12. Pu T., *et al.* Unlabeled Far-Field Deeply Subwavelength Topological Microscopy (DSTM). *Adv. Sci.* **8**, 2002886 (2020).
13. Wang M. C. & Uhlenbeck G. E. On the Theory of the Brownian Motion II. *Rev. Mod. Phys.* **17**, 323-342 (1945).
14. Berry M. V. & Shukla P. Hamiltonian curl forces. *Proc. R. Soc. A* **471**, 20150002 (2015).
15. Rodriguez A. W., Capasso F. & Johnson S. G. The Casimir effect in microstructured geometries. *Nat. Photon.* **5**, 211-221 (2011).

16. Aspelmeyer M. Gravitational quantum physics, or: How to avoid the appearance of the classical world in gravity experiments? 8th International Topical Meeting on Nanophotonics and Metamaterials; 2022 28-31 Mar. 2022; Seefeld-in -Tirol, Austria; 2022.
17. Liu T., *et al.* Ballistic Dynamics of Flexural Thermal Movements in a Nano-membrane Revealed with Subatomic Resolution. *Sci. Adv.* **8**, eabn8007 (2022).
18. Zhu W., *et al.* Quantum mechanical effects in plasmonic structures with subnanometre gaps. *Nat. Commun.* **7**, 11495 (2016).
19. Baumberg J. J., Aizpurua J., Mikkelsen M. H. & Smith D. R. Extreme nanophotonics from ultrathin metallic gaps. *Nat. Mater.* **18**, 668-678 (2019).
20. Yang B., *et al.* Sub-nanometre resolution in single-molecule photoluminescence imaging. *Nat. Photon.* **14**, 693-699 (2020).
21. Englander S. W. & Mayne L. The nature of protein folding pathways. *Proc. Natl. Acad. Sci.* **111**, 15873-15880 (2014).
22. Morozov S. V., *et al.* Giant Intrinsic Carrier Mobilities in Graphene and Its Bilayer. *Phys. Rev. Lett.* **100**, 016602 (2008).
23. Mariani E. & von Oppen F. Flexural Phonons in Free-Standing Graphene. *Phys. Rev. Lett.* **100**, 076801 (2008).
24. Lindsay L., Broido D. A. & Mingo N. Flexural phonons and thermal transport in graphene. *Phys. Rev. B* **82**, 115427 (2010).

# Supplementary Information:

# Picophotonics - Subatomic Optical Localization Beyond Thermal Fluctuations

Tongjun Liu[1]*, Cheng-Hung Chi[1], Jun-Yu Ou[1], Jie Xu[1], Eng Aik Chan[2], Kevin F. MacDonald[1], and Nikolay I. Zheludev[1, 2]

[1]Optoelectronics Research Centre and Centre for Photonic Metamaterials, University of Southampton; Highfield, Southampton, SO17 1BJ, UK.

[2]Centre for Disruptive Photonic Technologies, School of Physical and Mathematical Sciences and The Photonics Institute, Nanyang Technological University, Singapore; 637378, Singapore.

### S1: Wavefront synthesis

The computer-controlled wavefront synthesizer employed in this work is described in detail in Ref. 1. It is based upon a pair of (Meadowlark P512) spatial light modulators – one for intensity and the other for phase modulation.

For the purpose of this study we employ an axially-symmetric superoscillatory wavefront constructed from just two circular prolate spheroidal wavefunctions, *S3* and *S4* (following Rogers, *et al.*[2]): $E(r/\lambda)$ = 4.477 $S3(r/\lambda)$ + $S4(r/\lambda)$, where $r$ is radial distance from the beam axis. This simple analytical form considerably simplifies optimization of the experimental wavefront synthesizer, as only one free (relative weighting) parameter is involved.

In the 'plane wave' illumination regime, the synthesizer was configured to generate a defocused Gaussian beam profile having a (measured) intensity variance of only ±5% over the ~400 nm width of the sample (i.e. including the nanowire and gap on either side).

### S2: Neural network architecture, training, and application procedures

The neural network contained three convolution layers with, respectively, sixty-four 5×5, one hundred and twenty-eight 4×4, and two hundred and fifty-six 2×2 kernels, and three fully connected layers with 128, 256, 128 neurons. Each of the convolution layers was followed by a pooling layer with 4×4, 3×3, and 3×3 kernels with Rectified Linear Unit activation functions. The network was trained with the Adam stochastic optimization method and root mean square error loss function. The network was optimized by searching the hyperparameter space (adjusting the number of layers, number of neurons in each layer, and their activation functions) to simultaneously minimize training time and validation loss during training.

Our datasets comprise intensity patterns of transmitted light scattered by the nanowire at different electrostatically-controlled in-plane positions relative to the edges of the gap in the membrane. The patterns were imaged within a 10.3$\lambda$ × 10.3$\lambda$ (401 × 401 pixels) field of view of at a distance $\lambda$ from the sample. The scattering patterns were recorded at 301 different positions of the nanowire over a range from 0 to ~4.4 nm (applied bias settings of 0-2.1 V at intervals of 7 mV) in random sequence to excludes the possibility of neural network learning based upon any artefacts in the patterns other than those associated with changes in nanowire position. To eliminate any effect of stress history in the nanowire, its position was also reset to zero between each recorded position.

64% of scattering patterns (selected at random) were used for network training and 16% for validation, with the remaining 20% then employed for testing (i.e. as scattering patterns for nominally unknown nanowire positions, to be determined by the trained network). To exclude any dependence of measurement outcome on the selection of training scattering patterns and their order of appearance in the training process, twenty independent iterations of the training, validation and testing procedure were performed for each regime of illumination.

The recording of a complete set of 301 network training, validation and test images takes approximately 4 minutes. Over such a period, instrumental alignment fluctuations due to ambient mechanical noise and thermal instabilities (e.g. in the microscope frame, sample stage, etc.) may be orders of magnitude larger than the precision achieved in optical localization of the nanowire position relative to the slit edges. This is a strong indication that neural network training results in a retrieval algorithm which principally recognizes the structure of the scattering pattern created by the nanowire in the gap rather than its position on image sensor. Importantly, we find that measurement performance on such timescales is not substantively affected by whether test images are recorded among or immediately after training and validation images.

Ground truth values of nanowire displacement were independently established by *a priori* measurements under a scanning electron microscope for a number of different bias settings and interpolated by a quadratic dependence (Fig. S1): the first non-zero term in the analytical expression for the dependence of nanowire displacement ($D$) on applied bias ($V$) must be quadratic as displacement does not depend on the sign of the bias; and higher order terms are negligible while the magnitude of displacement remains much smaller that the gap size (of ~100 nm). Although each individual measurement by scanning electron microscope has an uncertainty of order ±1 nm (related to SEM image pixelation), measurements over a range of applied bias values enables accurate determination of the quadratic dependence $D = \alpha V^2$, where $\alpha$ takes a value of 1.0328 nm/V² with a standard error of 0.0053, in the present case. The resulting uncertainty in absolute calibration of displacement (applicable to both the actual and optically measured scales), at ~1.4%, is smaller than the achieved precision of optical measurements over the full measurement range.

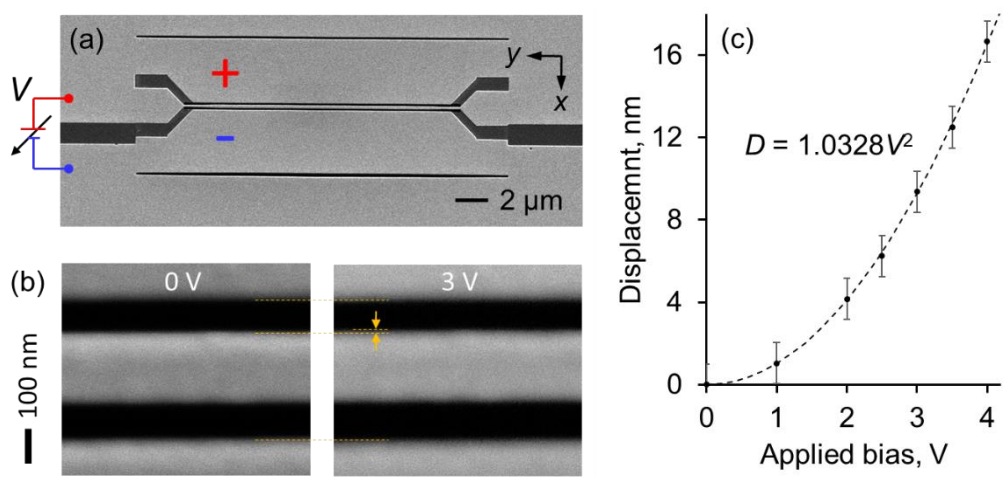

**Fig. S1. Nanowire position calibration**. (a) SEM image of the entire nanowire sample, showing the electrode configuration for electrostatic control of [*x* direction] position; (b) representative pair of high magnification images of the (*y* direction) midpoint of the nanowire from which the dependence of nanowire displacement on applied bias – panel (c) – is derived.

### S3: Numerical modelling of light scattering

Numerical simulations pertaining to the sensitivity of light scattering to small nanowire displacements (Fig. 3 in the manuscript) were performed using Lumerical FDTD Solutions. Silicon nitride is taken to have a refractive index *n* = (2 + 0*i*), while parameters for gold are those by Johnson & Christy. Incident light is polarized parallel to the nanowire and perfectly matching layer (PML) boundary conditions are used. The incident superoscillatory field was generated through a binary amplitude mask as detailed in Ref. 3.

**S4: Thermal fluctuations of the nanowire**

The thermal motion of nanomechanical structures is described by the Langevin model[4]. For a harmonic oscillator

$$\ddot{x} + \gamma\dot{x} + \omega_0^2 x = F_T(t)/m_{eff}$$

where

- $F_T(t) = \sqrt{2k_B T\gamma/m_{eff}}\eta(t)$ is the thermal force [which is related to the dissipation factor $\gamma$ through the fluctuation-dissipation theorem[5]];
- $k_B$ is the Boltzmann constant;
- $T$ is temperature;
- $\eta(t)$ is a delta-correlated normalized white noise term: $\langle\eta(t)\rangle = 0$; $\langle\eta(t)\eta(t')\rangle = \delta(t-t')$;
- $\omega_0 = 2\pi f_0 = \sqrt{(k/m_{\text{eff}})}$ is the natural angular frequency of oscillation, $f_0$ being the natural frequency and $k$ the spring constant;
- and $m_{eff}$ is the oscillator's effective mass;

the RMS beam displacement is

$$\delta x_{\text{RMS}} = \sqrt{k_B T/(4\pi^2 m_{\text{eff}} f_0^2)}$$

In the present case, $m_{eff}$ = 2 pg and for the in- and out-of-plane modes of oscillation respectively $f_0$ = 1.6 and 1.1 MHz, giving an average thermal fluctuation amplitudes of ~145 and ~215 pm.

**S5: Optical forces acting on the nanowire**

From numerical modelling of ponderomotive and radiation pressure forces, we conclude that within accuracy of the experiment they are insignificant to corrupt the ground truth values for the nanowire position.

To evaluate the action of these optical forces on the nanowire, we evaluated the Maxwell stress tensor in FDTD numerical simulations (Lumerical). We consider plane wave and superoscillatory incident wavefronts close to those used in experiment, with a total incident power of 100 µW over an 8 µm × 8 µm area of the sample. We then evaluate displacements induced by these forces, assuming a nanowire spring constant $k = m_{eff}\omega_0^2$ = 0.2 Nm⁻¹ (see section S4 above). In both illumination regimes, the optically-induced in-plane (*x* direction) displacement is zero when both the nanowire and incident light field are centered on the gap in the membrane, and it grows to not more than a few femtometers when the nanowire is laterally displaced (i.e. in experiment, electrostatically) by up to 5 nm. This represents a negligible perturbation against the tens of picometers best accuracy achieved in optical measurements of said displacement.

In the *z* direction perpendicular to the sample plane, optically-induced nanowire displacement (due to radiation pressure) may reach several tens of femtometers - still a very small amount. Moreover, the membrane on either side of the nanowire will be subject to comparable forces, whereby relative *z* displacement between nanowire and membrane will be much smaller, if not near-zero.

**References**

1. Rogers E. T. F.*, et al.* Far-field unlabeled super-resolution imaging with superoscillatory illumination. *APL Photon.* **5**, 066107 (2020).
2. Rogers K. S., Bourdakos K. N., Yuan G. H., Mahajan S. & Rogers E. T. F. Optimising superoscillatory spots for far-field super-resolution imaging. *Opt. Express* **26**, 8095-8112 (2018).
3. Yuan G., Rogers E. T. F. & Zheludev N. I. "Plasmonics" in free space: observation of giant wavevectors, vortices, and energy backflow in superoscillatory optical fields. *Light Sci. Appl.* **8**, 2 (2019).
4. Wang M. C. & Uhlenbeck G. E. On the Theory of the Brownian Motion II. *Rev. Mod. Phys.* **17**, 323-342 (1945).
5. Kubo R., Toda M. & Hashitsume N. *Statistical Physics II: Nonequilibrium Statistical Mechanics*, vol. 31. Springer: Berlin, 1991.